\input harvmac
\overfullrule=0pt
\input epsf.tex

\Title{}
{\vbox{\centerline{Resolving the black hole information paradox*
 }}}
\smallskip
\smallskip
\centerline{\bf Samir D. Mathur}
\smallskip
\bigskip

\centerline{\it Department of Physics}
\centerline{\it The Ohio State University}
\centerline{\it Columbus, OH 43210, USA}
\medskip
\centerline{mathur@mps.ohio-state.edu}
\bigskip

\medskip

\noindent

The recent progress in string theory strongly suggests that formation and
evaporation of black holes is a unitary process. This fact makes it imperative that we
find a flaw in the semiclassical reasoning that implies a loss of information.  We propose
a new criterion that limits the domain of classical gravity: the hypersurfaces of a
foliation cannot be {\it stretched} too much. This conjectured criterion may have
important consequences for the  early Universe. 

\vskip 2.0 true in

*This essay received an ``honorable mention'' in the 
	Annual Essay Competition of the Gravity Research 
	Foundation for the year 2000. 

\Date{}

It is intriguing that black holes behave as if they had an entropy proportional to
the area of their horizon \ref\bek{J.D.  Bekenstein: 1973, 
{\it Phys. Rev.} {\bf D7}, 
 2333.}, and the discovery of Hawking that black holes radiate 
thermally accorded beautifully with the requirements of  thermodynamics
\ref\haw{S. Hawking: 1975,  {\it Comm. Math. Phys.} {\bf 43}, 
 199.}. But
the calculation of this radiation raised a sharp paradox. The emerging particles have
little information about the matter that made the  hole; worse they are in an
 entangled state with their antiparticles that fall into the hole and decrease
its mass. If the hole evaporates away then an initial pure state evolves to a mixed
state, leading Hawking to postulate that in the presence of gravity we must generalize
quantum mechanics so that systems  are described not by
wavefunctions but by density matrices. 

The recent progress in string theory, however, suggests very strongly that black
holes behave just like any other composite particle in quantum mechanics: their
entropy can be attributed to the degeneracy of their internal states, and Hawking
emission is a unitary process where the de-excitation of an internal state leads to 
emission of a quantum.   In particular the entropy of extremal \ref\sv{A. Strominger and C. Vafa: 1996, 
{\it Phys. Lett.} {\bf B379}, 
 99.} and near
extermal \ref\cm{C. Callan and J. Maldacena: 1996, 
{\it Nucl. Phys.} {\bf B472}, 
 591.} black holes have been reproduced exactly from a count of microstates. The
spin dependence and radiation rate of the low energy Hawking radiation is exactly
reproduced from the dynamics of these microstates \ref\dm{S.R. Das and S.D. Mathur: 1996, 
{\it Nucl. Phys.} {\bf B478}, 
561.}, and in a certain near
extremal limit one finds that  greybody factors are
reproduced as well \ref\ms{J. Maldacena and A.
Strominger:  1997 {\it Phys. Rev} {\bf D55},   861.}. 

If we accept these suggestions from string theory, then we solve the information {\it
problem}: the information in the matter  thrown into the hole is recovered in
the radiation that emerges, and there is no need for a modification of quantum
mechanics. But it does not solve the information {\it paradox}: as with any paradox,
we are presented with an explicit argument (in this case the semiclassical calculation
of Hawking) and are then challenged to point out which step in the argument is
fallacious.  The string theory results require extrapolations from weak coupling
(invoking the invariances guaranteed by supersymmetry), and thus do not give an
explicit picture of information recovery. The AdS/CFT duality \ref\malda{J. Maldacena: 1998, 
{\it Adv. Theor. Math. Phys.} {\bf 2}, 
 231,  S. Gubser, I. Klebanov and A. Polyakov: 1998, 
{\it Phys. Lett.} {\bf B428}, 
 105,  E. Witten: 1998, 
{\it Adv. Theor. Math. Phys.} {\bf 2}, 
 253.
} does not do
any better, since it is well formulated only for `global AdS' which has no horizons; in
any event the dual gravity theory is understood only as perturbative strings and
thus offers no hint of how infalling information will be redirected out.

The paradox can be formulated as follows.  One can foliate the black hole spacetime such
that (a) all slices (spacelike hypersurfaces) are intrinsically smooth (b) the embedding
of these hypersurfaces in the spacetime is smooth  (c)  all slices capture the matter that fell in to make
the hole; the late time slices also capture the Hawking radiation near spatial infinity and their
corresponding antiparticles inside the horizon.\foot{We also assume that the slicing of spacetime is {\it
stable}, in the sense that a small change in the intrinsic geometry of the slice does not lead to a large
change in the state of the matter on the slice. Slicings that were not stable were investigated 
in \ref\eskoone{E.  Keski-Vakkuri, G. Lifschytz, S. D. Mathur and M. Ortiz: 1995,
{\it  Phys.Rev.} {\bf D51}, 1764. }, but it was argued
 in \ref\eskotwo{E. Keski-Vakkuri  and  S. D. Mathur: 1996, {\it  Phys.Rev.} {\bf D54},7391. }
that consideration of unstable slices did not affect the information question.}

Since we cannot have  `quantum xeroxing' we must have
somehow `bleached' the information from the initial matter as seen on this final
surface, and transferred its information to the radiation that appears on this same
surface near infinity. But we cannot simply postulate a new nonlocal mechanism of
information transfer: since the entire evolution appears to be within the domain of
semiclassical physics, how do we ensure that we still recover `normal physics' in the
absence of black holes, where such information transfer is not observed to occur?
 What we need therefore is a
{\it criterion} for when this nonlocal information transport will be triggered: the
criterion must cover the black hole evaporation process but not ordinary everyday
evolution.\foot{The idea that nonlocal effects take place in in the context of black holes is not itself new
(see for example \ref\ah{D. V. Ahluwalia: 1994, {\it Phys. Lett.} {\bf B339}, 301.}), but what we seek
here is a way to bypass the Hawking argument for information loss while preserving our experience of
low energy physics.}

We propose that a fourth criterion (d) needs to be added to the above
three, to define a domain where semiclassical intuition should be valid. Consider a
`sandwich' of spacetime formed by an initial spacelike slice, a final spacelike slice, and
the spacetime in between.  We postulate that a slice in spacetime needs to be specified
not only by its intrinsic and extrinsic geometries, but also by a {\it density of degrees
of freedom}, defined over the slice.  We require that the count of these degrees of
freedom is conserved. If the evolution through the `sandwich' causes the slice to
`stretch', then the final slice will have a lower density of degrees of freedom. We then
require that usual semiclassical physics holds as long as we do not load any region of a slice
with more bits of data than there are degrees of freedom available in that region.  If
on the other hand we do try to overload a region on a slice with more data than there
are bits, then degrees of freedom must be recruited from outside the region, and
nonlocal information transport is possible.

It turns out that in the process of Hawking
radiation whenever we try to choose a foliation that satisfies criteria (a), (b), (c)
above, then we are forced to loss of classicality  through criterion (d); thus
 Hawking's semiclassical derivation  (which used classical geometry and quantum
matter) need not hold after a brief initial period of evolution that we will estimate. But
this outcome would appear to create trouble with the other crucial issue with the
 paradox: based on the equivalence principle, we expect that an infalling
observer should see nothing special as he enters the black hole horizon and moves
towards the singularity. If we have a nonclassical spacetime inside the hole, will the
observer not feel his motion to be nonclassical inside the hole?

The resolution of this problem which we  propose is the following.  An infalling
quantum must have an energy greater than the temperature of the hole, in order
that it be described by a classical path. But in this case its energy would bring into the
hole a sufficient number of additional degrees of freedom, so that the particle can
`ride' over the spacetime formed by these degrees of freedom, for the duration
required to describe its motion from the horizon to the vicinity of the singularity. This
picture is suggested by the explicit solution of 1+1 dimensional  string theory, 
where excitations are mapped to distortions of the fermi surface of a free fermion
liquid. Even though an ingoing pulse may `spill over' and form a black hole, further
incoming quanta are formed out of their own collection of fermions, and (seen in a
certain description) travel without distortion \ref\pn{M. Natsuume and J. Polchinski: 1994
{\it Nucl. Phys.} {\bf 424}, 137. }\ref\dmpre{S. Das and S.D. Mathur: 1996, {\it Phys.
Lett} {\bf B365},   79.}.

If the foliation violates the criteria (a)-(c) then one is not forced to a paradox: some
part of the slices would  have strong curvature  and new physics can be
claimed to occur and cause nonlocal information transport. What we note here is that if
we do satisfy (a)-(c) in foliating the black hole geometry, then we will necessarily find
that late time slices are highly `stretched' compared to the initial slices, even though
the evolution from any one slice to a neighboring one exhibits no unusual features.
Rather than prove this rigorously we sketch in the figure a foliation\foot{This foliation is essentially
similar to one used for example in \ref\suss{D.  Lowe, J. Polchinski, L. Susskind, L. 
Thorlacius and  J. Uglum: 1995,  {\it  Phys.Rev.} {\bf  D52}, 6997.} } satisfying (a)-(c). The
figure is not a Penrose diagram; it is just a plot of the
$r,t$ coordinates of the Schwarzschild metric
\eqn\one{ds^2=(1-{2M\over r})~dt^2+( 1-{2M\over
r})^{-1}dr^2+r^2(d\theta^2+\sin^2\theta d\phi^2)}
We draw two spacelike slices, ABCD and a later one ABB'C'D'. The slices are constant $t$
near infinity, and go over to constant $r$ inside the hole (where constant $r$ is
spacelike).  We label the slices by $t$, and let $t_2-t_1\sim M$.

Assuming that the degrees of freedom follow the normal to the slice in the embedding
manifold, we see that the degrees in CD move to C'D', while those in the part AB stay
frozen. But those in BC must stretch to cover BB' as well as B'C'. 
Thus the density in B'C' will be lower  than in BC by some factor
$\alpha<1$.  As we continue the evolution, we find
that there is a progressive dilution of the degrees of freedom in regions like B'C',
which turn out to have a density  $\sim e^{-\alpha t/M}$. 
The geometry will cease to be classical when
\eqn\two{S_{BH}e^{-\alpha{t\over M}}~<~f}
Here $S_{BH}= A/(4\pi G)$ estimates the degrees of freedom present on the initial slice in the balck hole
region, and  $f$ is the number of matter fields that are radiated (the negative energy antiparticles need
$f$ bits for each interval of length $M$ on the slice).
  Thus by our new criterion the
foliation will cease to be classical at quite an early time $t\sim M\log M$ (for $f\sim 1$), and so after this time
there is no contradiction with semiclassical physics if we allow the Hawking quanta to carry the
information of the initial matter, 
nonlocally transported across  nonclassical regions like
 B'C'. Of course if there is no such `overstretching' of slices in an evolution, we will continue to have
`normal' physics, and the paradox is thus resolved.

 Now consider an infalling quantum, with energy $\delta
E>>T_H$. The increase in the black hole entropy upon absorbing this energy is 
\eqn\three{\delta S\sim {dS\over dE}\delta E \sim {1\over T_H}\delta E>>1}
Thus the number of states after absorbing the quantum $\sim e^{S+\delta S}$ is much
higher than the number of states $\sim e^S$ that went to make up the hole. These new
states can describe a classical spacetime for the duration where the infalling quantum
falls through a distance $\sim M$ in the hole, after which they will
also be redistributed in the ever stretching slices required by the nature of the
foliation.  Thus there is no contradiction between the fact that the foliation becomes nonclassical, and the
requirement that an infalling object experience classical motion through the horizon.

The essence of a physical postulate is of course  that it must be applied to
all physical systems. In particular if we consider an  expanding Universe then, given
our postulate, we must start with some given density of degrees of freedom on an
initial slice, and see this density drop as the expansion proceeds. Thus we should not
be able to `load' as much matter on a unit volume of space today as we could in the
past when this density was higher.  If the expansion proceeds to a point where there
are less degrees of freedom than that required to describe the matter content, then
nonlocal information transport would be allowed, with important consequences for the
question of homogeneity beyond the Hubble scale. Further, if we trust the Bekenstein
relation $S=A/(4\pi G)$ as giving some measure of degrees of freedom in any region
\ref\fs{W. Fischler and L. Susskind, hep-th/9806039.}, then a lower density of these
degrees would correspond to a higher value of the
 $G$; we would  find a time varying Newton's constant with corresponding
changes in the geometry of the initial singularity. Clearly, it is imperative that we
extract the lessons of the black hole information paradox so that we may understand
the structure of spacetime in diverse situations!

\vfill
\eject
\epsfbox{  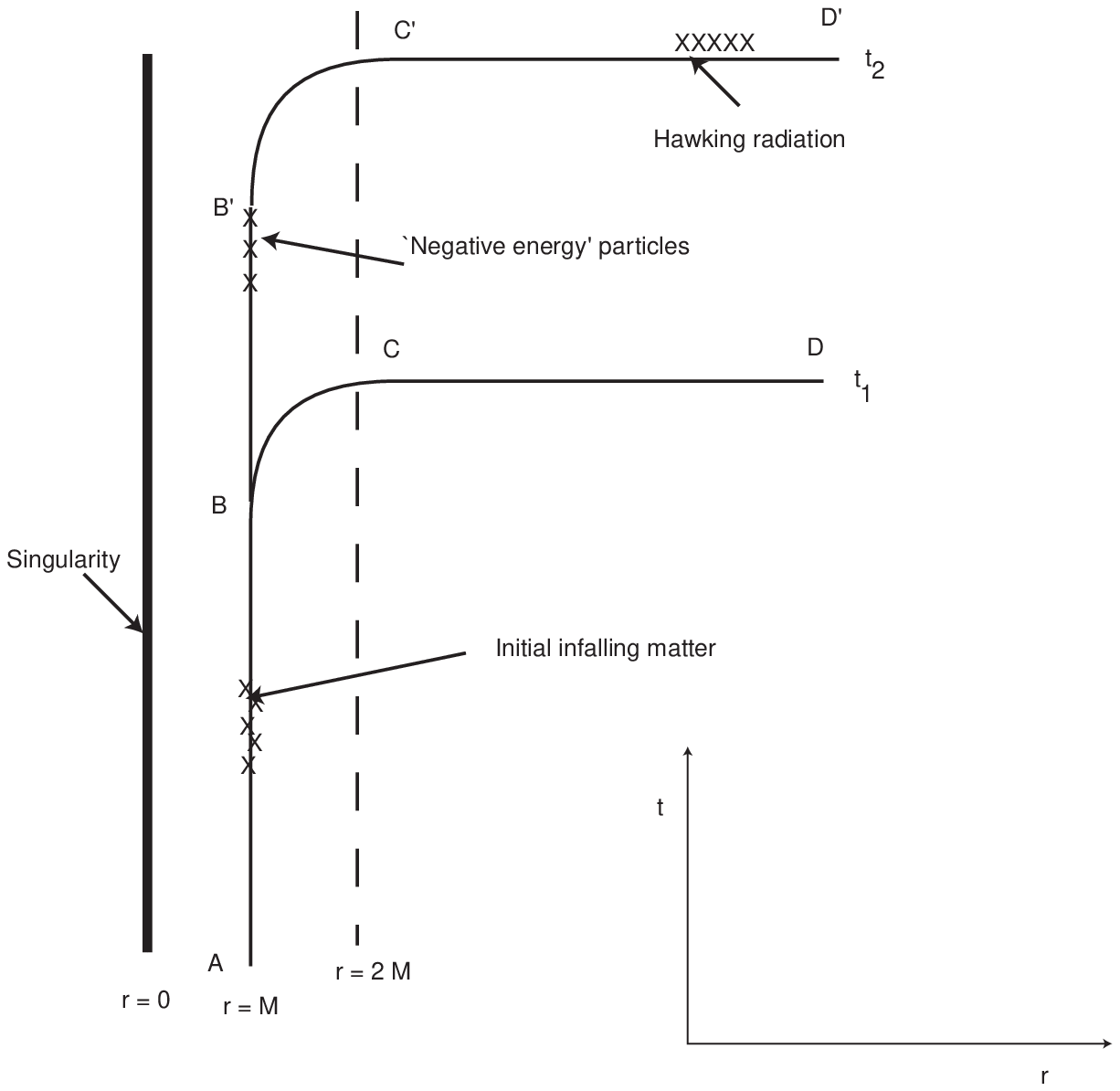} 
\eject
\listrefs

\bye

\end